# Antenna-coupled silicon-organic hybrid integrated photonic crystal modulator for broadband electromagnetic wave detection

Xingyu Zhang*[,a], Amir Hosseini[b], Harish Subbaraman[b], Shiyi Wang[c], Qiwen Zhan[c], Jingdong Luo[d], Alex K.-Y. Jen[d], Chi-jui Chung[a], Hai Yan[a], Zeyu Pan[a], Robert L. Nelson[e], Charles Y.-C. Lee[e], and Ray T. Chen*[,a,b]

[a] University of Texas at Austin, 10100 Burnet Rd, MER 160, Austin, TX 78758, USA;
[b] Omega Optics, Inc., 8500 Shoal Creek Blvd, Austin, TX 78757, USA;
[c] University of Dayton, 300 College Park, Dayton, OH 45469-2951, USA;
[d] University of Washington, 302 Roberts Hall, Seattle, Washington 98195, USA;
[e] Air Force Research Laboratory at Wright Patterson, Dayton, Ohio 45433, USA.

## ABSTRACT

The detection and measurement of electromagnetic fields have attracted significant amounts of attention in recent years. Traditional electronic electromagnetic field sensors use large active conductive probes which perturb the field to be measured and also make the devices bulky. In order to address these problems, integrated photonic electromagnetic field sensors have been developed, in which an optical signal is modulated by an RF signal collected by a miniaturized antenna. In this work, we design, fabricate and characterize a compact, broadband and highly sensitive integrated photonic electromagnetic field sensor based on a silicon-organic hybrid modulator driven by a bowtie antenna. The large electro-optic (EO) coefficient of organic polymer, the slow-light effects in the silicon slot photonic crystal waveguide (PCW), and the broadband field enhancement provided by the bowtie antenna, are all combined to enhance the interaction of microwaves and optical waves, enabling a high EO modulation efficiency and thus a high sensitivity. The modulator is experimentally demonstrated with a record-high effective in-device EO modulation efficiency of $r_{33}$=1230pm/V. Modulation response up to 40GHz is measured, with a 3-dB bandwidth of 11GHz. The slot PCW has an interaction length of 300 μm, and the bowtie antenna has an area smaller than 1cm$^2$. The bowtie antenna in the device is experimentally demonstrated to have a broadband characteristics with a central resonance frequency of 10GHz, as well as a large beam width which enables the detection of electromagnetic waves from a large range of incident angles. The sensor is experimentally demonstrated with a minimum detectable electromagnetic power density of 8.4mW/m$^2$ at 8.4GHz, corresponding to a minimum detectable electric field of 2.5V/m and an ultra-high sensitivity of 0.000027V/m Hz$^{-1/2}$ ever demonstrated. To the best of our knowledge, this is the first silicon-organic hybrid device and also the first PCW device used for the photonic detection of electromagnetic waves. Finally, we propose some future work, including a Teraherz wave sensor based on antenna-coupled electro-optic polymer filled plasmonic slot waveguide, as well as a fully packaged and tailgated device.

## 1. INTRODUCTION

The detection, measurement and evaluation of electromagnetic fields have attracted a significant amount of attention in recent years [1-4]. Electromagnetic field sensors have shown promising applications in high power microwave (HPM) detection, electromagnetic pulse (EMP) detection, environmental electromagnetic interference (EMI) analysis, electromagnetic compatibility (EMC) measurements, radio frequency (RF) integrated circuit testing, process monitoring and control, as well as in the research of electromagnetic radiation effects on human health. Traditional electronic-based electromagnetic field sensors [5, 6] normally have large active conductive probes which always perturb the field to be measured and also make the device bulky. In order to address these problems, integrated photonic sensing of electromagnetic field has been developed, in which the optical signal is modulated by an RF signal collected by an antenna [7-12]. The antennas used here can potentially be designed to be small enough or all-dielectric [13] to minimize the perturbation of high-frequency electric field under measurement. The key element of such devices is an efficient electro-optic (EO) modulator. One common structure of EO modulators is a Mach-Zehnder interferometer (MZI), in which an electric-field-induced optical phase modulation is converted into an optical intensity variation [14, 15]. These integrated photonic electromagnetic field sensors have a few inherent advantages over conventional electronic sensors, including compact size, high sensitivity, broad bandwidth, good galvanic insulation and noise immunity [16].

*xzhang@utexas.edu; phone 1 512-471-4349; fax 1 512 471-8575
*raychen@uts.cc.utexas.edu; phone 1 512-471-7035; fax 1 512 471-8575

Recently, silicon-organic hybrid (SOH) technology [17] has shown to enable high performance integrated photonic devices such as compact and low-power EO modulators [18, 19], high-speed optical interconnects [20, 21], and sensitive photonic sensors [22, 23]. Benefiting from the large EO coefficient ($r_{33}$) of active organic polymers [24, 25] as well as the strong optical mode confinement made possible by the large index of silicon [26, 27], SOH integrated photonic electromagnetic field sensors are promising for achieving high sensitivity, compact size, and broad bandwidth.

It has been demonstrated in Ref. [16] that low half-wave voltage ($V_\pi$), high input driving voltage (electric field), and large input optical power of EO modulators can improve the sensitivity of the photonic electromagnetic field sensors. To enhance the sensitivity through $V_\pi$ reduction, silicon slot photonic crystal waveguides (PCWs) refilled with large EO coefficient ($r_{33}$) EO polymer can be employed due to the slow-light enhanced light-matter interaction [28, 29]. The enhanced light-matter interaction also enables a compact device size [30, 31]. The silicon in the SOH slot PCW modulator can also be doped to achieve high speed operation [32, 33]. For example, modulation speeds up to 40GHz or over 40Gbit/s have been demonstrated in [34, 35]. Furthermore, the antenna can be designed for broadband resonant electric field enhancement [36, 37], which is equivalent to increasing the input driving voltage of the modulator, thereby increasing the sensitivity of the sensor. The geometrical dimensions of the antenna are much smaller than the wavelength of the electromagnetic field to be measured. This antenna can be combined with the SOH slot PCW modulator to achieve even higher sensitivity over wide frequency bandwidth. Such a combination of integrated RF photonics [38] and SOH technology [17] offers a viable platform for high frequency electromagnetic field sensing.

In this paper, we design and demonstrate an integrated photonic electromagnetic field sensor based on bowtie antenna coupled SOH slot PCW modulator. The modulator is experimentally demonstrated with a record-high effective in-device EO modulation efficiency of $r_{33}$=1230pm/V. Modulation response up to 40GHz is measured, with a 3-dB bandwidth of 11GHz. The measured S-parameters show that the bowtie antenna with slot PCW embedded inside its feed gap has broadband characteristic, with a resonance peak at 10GHz. The measured radiation pattern of the bowtie antenna indicates large beam width which promises the detection of electromagnetic waves from a large range of incident angles. High frequency electromagnetic field sensing is experimentally demonstrated through EO modulation at 8.4GHz, with a minimum detectable electromagnetic power density of 8.4mW/m$^2$, corresponding to an incident electric field as small as 2.5V/m and an ultra-high sensitivity of 0.000027V/m Hz$^{-1/2}$ ever demonstrated [14]. To the best of our knowledge, this is the first silicon-organic hybrid device and also the first PCW device used for the photonic detection of electromagnetic waves.

## 2. DESIGN

### A. Device overview

The key parts of our photonic electromagnetic field sensor consisting of an EO polymer refilled silicon slot PCW phase modulator coupled with a gold bowtie-shaped antenna are shown schematically in Figs. 1 (a)-(e). The slot and holes of the silicon PCWs are filled with an EO polymer (SEO125 from Soluxra, LLC), which has an exceptional combination of large EO coefficient ($r_{33}$ of ~100pm/V at 1550nm), low optical loss, synthetic scalability, as well as excellent thermal- and photochemical-stability [39]. The refractive index of the EO polymer can be changed by applying an electric field via the EO effect (also called Pockels effect). The slow-light effect in the PCW can enhance the interaction of RF field and optical waves and thus increases the effective in-device EO coefficient of this SOH modulator to be a value larger than 1000pm/V [19, 39], which is beneficial for high-sensitivity sensing. The silicon slot PCW is embedded in the feed gap of the bowtie antenna, and the silicon layer is selectively implanted with different ion concentrations for high frequency operation [40]. The bowtie antenna is used as a receiving antenna, driving electrodes, and poling electrodes. Here the bowtie antenna with capacitive extension bars has a simple design, and a broadband characteristic. With the two bowtie arms as receivers, a confined resonant electric field with strong enhancement factor can be generated in the gap [36].

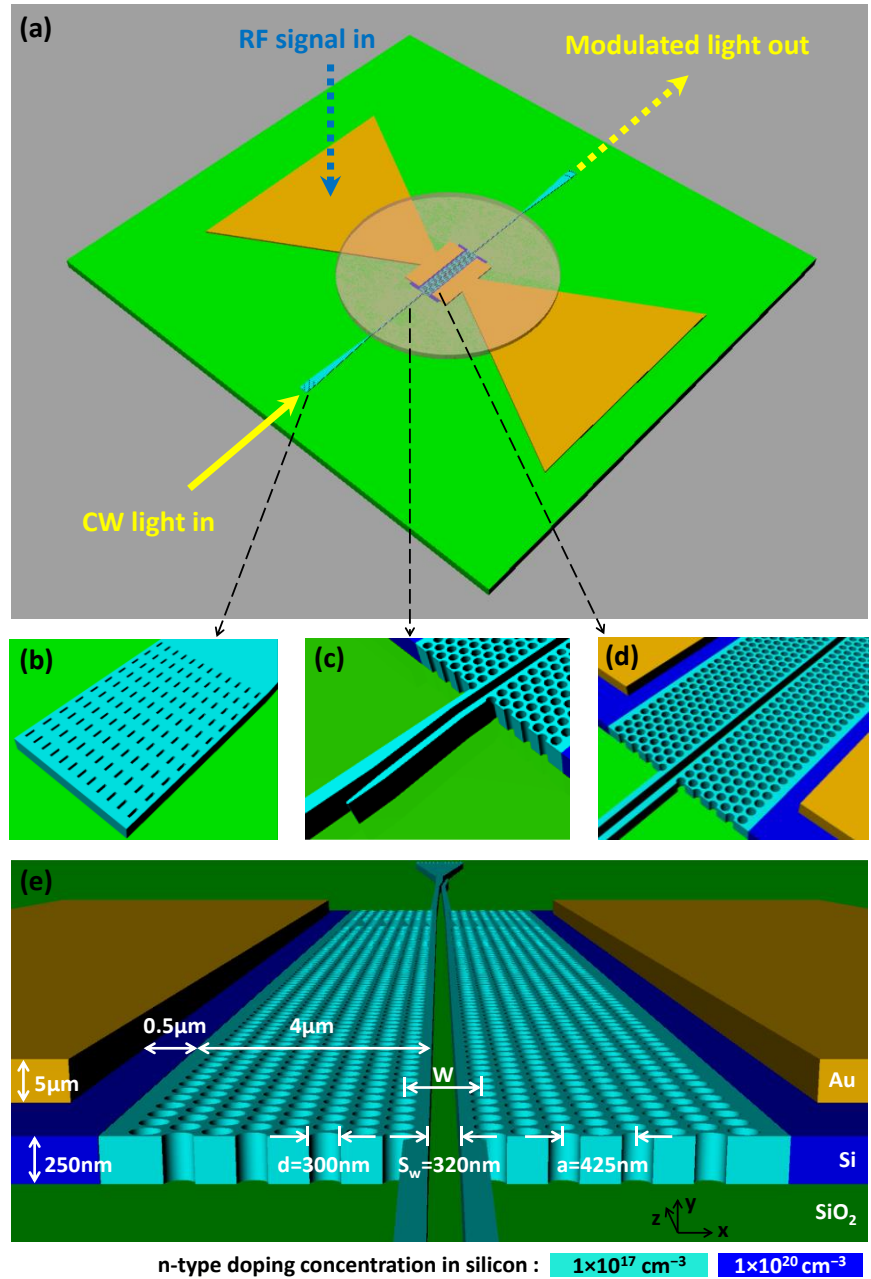

Fig. 1. (a) A schematic view of the key part of the electromagnetic field sensor consisting of an EO polymer refilled silicon slot PCW phase modulator and a bowtie antenna. An external arm combined with this phase modulator forms an MZI structure, converting phase modulation to intensity modulation. (b) Subwavelength grating coupler. (c) Adiabatic strip-to-slot mode converter (d) Magnified image of slot PCW. (e) Tilted view showing the cross section of the antenna-coupled slot PCW, with dimension parameters and two levels of n-type silicon doping concentrations. Note: the EO polymer layer covered on top of the device is not shown in (b)-(e) for better visualization.

The working principle of this integrated photonic electromagnetic field sensor is discussed as follows. A continuous wave (CW) laser input is coupled into and out of the device through subwavelength grating (SWG) couplers [Fig. 1 (b)] [41, 42]. The bowtie antenna harvests incident electromagnetic waves, transforms it into high-power-density time-varying electric field within the feed gap, which directly interacts with the light propagating along the EO polymer refilled slot PCW embedded within the feed gap (interaction region). The refractive index of the EO polymer is controlled by the applied electric field via the EO effect, which modulates the phase of the optical wave. For measurement, we convert this phase modulation to an intensity modulation using an external arm enabled MZI structure. Finally, by measuring the modulated optical intensity at the output end of the MZI, an incident electromagnetic field from free space can be detected through optical means. This integrated photonic electromagnetic field sensor can reduce the impact of perturbing fields, since it is based on an optical modulation technique. The sensitivity of the device is strongly enhanced by three combined factors: (1) the large $r_{33}$ of the EO polymer material, (2) the slow-light effect in the silicon PCW, and (3) the field enhancement provided by the gold bowtie antenna.

**B. Design of silicon-organic hybrid slot photonic crystal waveguide modulator**

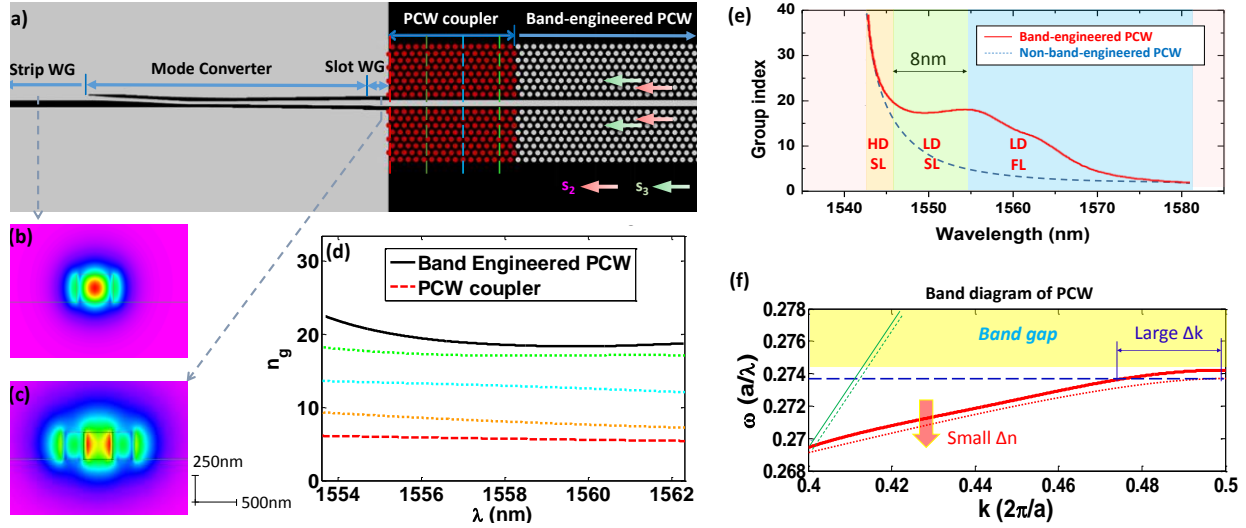

Fig. 2. (a) Layout of the strip-to-slot mode converter, PCW coupler and band-engineered slot PCW. The black area corresponds to un-etched silicon, and the gray area corresponds to the etched silicon. The red-color holes indicate the PCW taper section. The $s_1$, $s_2$, and $s_3$ indicate the lattice shift direction in band-engineered PCW section. WG: waveguide. (b) Cross-sectional view of strip waveguide mode profile. (c) Cross-sectional view of slot waveguide mode profile. (d) The small variations of $n_g$ over about 8nm wavelength range, for the band-engineered slow-light PCW and PCW coupler. The lines of different colors represent the $n_g$ at different positions along the PCW coupler as indicated by the dashed lines of corresponding colors in (a). This indicates a smooth transition of $n_g$ from the beginning of the PCW taper to the band-engineered PCW. (e) A band diagram showing the simulated dispersion curve of the band-engineered EO polymer refilled slot PCW (solid red curve), overlaid with that of a regular strip waveguide (green curve). The small change of EO polymer index leads to a larger change of k vector (thus larger phase shift) in the slot PCW than in the regular strip waveguide. (f) Simulation result of engineered group index in the slot PCW (red curve) as a function of wavelength, showing 8nm low-dispersion slow-light wavelength region (flat band nature of low-dispersion region highlighted in green). Also overlaid is a blue dashed curve representing the dispersive group index versus wavelength for non-band-engineered PCW for comparison. HD SL: high-dispersion slow-light; LD SL: low-dispersion slow-light; LD FL: low-dispersion fast-light.

The layout of the slot PCW is shown in Fig. 2 (a). This slot PCW is designed on a silicon-on-insulator (SOI) substrate with 250nm-thick top silicon and 3 μm-thick buried oxide layers. The PCW holes and slot are assumed to be filled with EO polymer, SEO125, which has a refractive index of 1.63 at 1550nm. The slot width in the PCW is designed to be $S_w$=320nm, which not only supports a confined optical mode but also helps in increasing the poling efficiency by suppressing the leakage current in EO polymer poling process [39, 43]. A lattice-shifted slot PCW is designed to achieve low-dispersion slow light [44], where optimized values of the lattice constant (a=425nm), hole diameter (d=300nm), center-to-center distance between two rows adjacent to the slot (W=1.54($\sqrt{3}$)a), relative lateral shift of the first three rows along the slot ($s_1$=0, $s_2$=-85nm, $s_3$=85nm [indicated by the arrows in Fig. 2 (a)]) are utilized. A group index ($n_g$) is engineered to be almost constant value of 20.4 (±10%) over a broad wavelength range of 8nm, in order to provide low-dispersion slow-light enhancement [39]. The $n_g$ of this band-engineered PCW simulated by Rsoft BandSolve is shown by the red solid curve in Fig. 2 (e). The 8nm-wide low-dispersion spectrum enables a relatively larger operational optical bandwidth of device, compared to non-band-engineered PCWs [blue dashed curve in Fig. 2 (e)] with typically narrow optical bandwidth of <1nm at $n_g$>10 [45], and it also makes our device less sensitive to the variation of temperature and wavelength.

In order to efficiently couple light from a strip waveguide into the 320nm-wide-slot waveguide, an adiabatic strip-to-slot mode converter [46], as shown in Fig. 1 (c), is used to serve as a smooth transition from the strip mode to the slot mode [Figs. 2 (b) and (c)]. In addition, to reduce the coupling loss between the fast-light mode in the slot waveguide at the mode converter end ($n_g$~3) and the slow-light mode of the slot PCW ($n_g$~20.4), a PCW taper consisting of non-band-engineered PCW (a=425nm, d=300nm, $s_1$=0, $s_2$=0, $s_3$=0, $S_w$=320nm) is designed and inserted between the two device components, as indicated in Fig. 2 (a), in which W increases parabolically from W=1.45($\sqrt{3}$)a to W=1.54($\sqrt{3}$)a from the beginning to the end of the PCW taper. Fig. 2 (d) shows the $n_g$ variation of the PCW taper over about 8nm wavelength range for the engineered slow-light PCW and the PCW coupler. The colored dashed lines show the gradual increase in the group index from the interface with the mode converter to the interface with the high $n_g$ band-engineered PCW.

The photonic band diagram in Fig. 2 (f) shows the simulated dispersion curve of the band-engineered EO polymer refilled slot PCW, as indicated by the solid red curve, in which the slope indicate the $n_g$ value. It can be seen that the group index (group velocity) close to the bandwidth is very large (slow). A small variation of EO polymer index (($\Delta n$) can cause a slight vertical shift of the dispersion curve, leading to a large change of k vector and thus a large change of phase shift ($\Delta\varphi=\Delta k \times L$). This explains how slow-light effect in the PCW plays its role in enhancing the phase modulation efficiency. In comparison, a regular strip waveguide, indicated by the green curves in Fig. 2 (f), cannot provide such a large $\Delta\varphi$ under the same $\Delta n$.

The basic sensing principle is based on the EO modulation inside the EO polymer refilled silicon slot PCWs. The required PCW interaction length for achieving a $\pi$ phase shift is given as $L=1/(2\sigma)\times(n/\Delta n)\times\lambda/n_g=255.9\,\mu m$, where $\sigma=0.33$ is the fraction of the energy in the slot calculated using BandSolve simulations, n=1.63 is the index of the EO polymer, $\Delta n$=0.0007 is the change in the index of the EO polymer when voltage V=1V is applied, and $\lambda$=1550nm is the free-space wavelength and $n_g$=20.4 is the group index. The change in the EO polymer index is calculated using $\Delta n=-n^3 r_{33}V/(2S_w)$, where the estimated $r_{33}$=100pm/V at 1550nm is consistent with the large $r_{33}$ value of 125pm/V of SEO125 thin films at 1.3μm after considering the dispersion factor and the nearly 100% poling efficiency demonstrated in 320nm-wide slots [39]. Therefore, from these calculations, the figure of merit of the modulator is $V_\pi \times L$=1V×255.9μm=0.0256V×cm. Such a small $V_\pi \times L$ promises a compact and efficient EO modulator, and thus a highly sensitive integrated photonic electromagnetic field sensor. The expected effective in-device $r_{33}$ is then calculated as [47]

$$r_{33,effective} = \frac{\lambda S_w}{n^3 V_\pi \sigma L} = 1356 \text{pm/V} \qquad (1)$$

In the Section 4 – Characterization, we are able to experimentally demonstrated the $V_\pi \times L$ and effective in-device $r_{33}$ on the same order [39]. Conservatively, the length of the active PCW section used in our work is chosen to be 300μm.

## C. Design of RF bowtie antenna

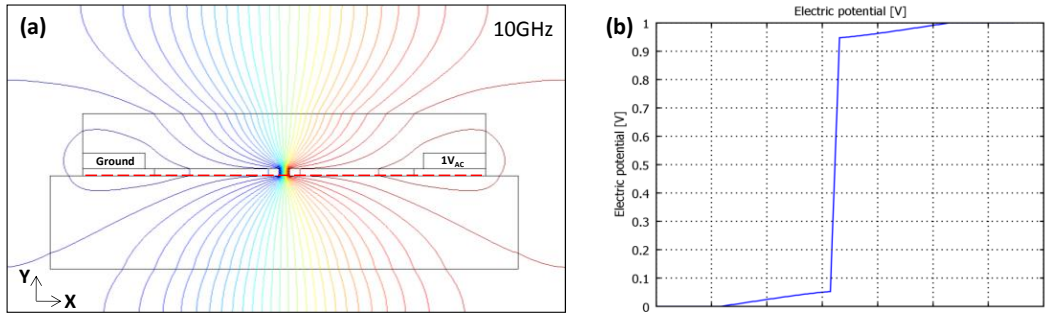

Fig. 3. (a) Cross-sectional view of RF (10GHz) electric potential distribution across the EO polymer refilled doped silicon slot PCW. (b) Electric potential along the red dashed line in (a), indicating a large percent of voltage dropped across the slot.

To achieve modulation (or sensing) in GHz frequency regime, both the silicon PCW layer and the bowtie antenna need to be specially designed and carefully optimized. It is known that the RC time delay of a lumped device is one key factor limiting its RF bandwidth. In order for our sensor to operate at a high frequency, the top silicon layer is doped to reduce the electrical resistivity of the PCW, while maximizing the electric field inside the slot. The designed two-level doping condition in the silicon slot PCW is shown in Fig. 1 (e). The resistivity values of highly-doped silicon (donor, $1\times10^{20} cm^{-3}$) and low-doped (donor, $1\times10^{17}cm^{-3}$) are $9\times10^{-6}\Omega\cdot m$ and $9\times10^{-4}\Omega\cdot m$, respectively [48]. For reference, the intrinsic doping concentration of the undoped top silicon on our SOI wafer is $1\times10^{14}cm^{-3}$. Note that an ion doping concentration of $1\times10^{17}cm^{-3}$ in the waveguide region close to optical mode does not cause significant impurity-induced scattering optical loss [49]. Based on our previous work [18, 43, 50, 51], in the case of 320nm-wide slots, we use the EO polymer resistivity ($\rho_{EO}$) value of about $10^{8}\Omega\cdot m$ and RF dielectric constant ($\varepsilon_{RF,EO}$) value of 3.2. The change in the RF dielectric constant of silicon ($\varepsilon_{RF,Si}$) due to the doping is also taken into account [52]. Then, effective medium approximations [53] are used for the calculation of both the effective RF dielectric constant ($\varepsilon_{RF,eff}$) and the effective resistivity ($\rho_{eff}$) in the region of EO polymer refilled silicon PCW shown in Fig. 1 (e) [hexagonal lattice, filling factor (volume fraction): f=0.444]. The effective RF dielectric constant (electric field in x-direction) in this region is given as [54].

$$\varepsilon_{RF,eff} = \varepsilon_{RF,Si}\left[\frac{\varepsilon_{RF,EO}(1+f)+\varepsilon_{RF,Si}(1-f)}{\varepsilon_{RF,EO}(1-f)+\varepsilon_{RF,Si}(1+f)}\right] \quad (2)$$

The effective resistivity in this region is estimated as [55]

$$\rho_{eff} = \rho_{Si}\left(\frac{1+f}{1-f}\right) \quad (3)$$

where, $\rho_{Si}$ is the resistivity of un-patterned silicon.

Fig. 3 (a) shows the cross-sectional view of the RF electric potential distribution ($1V_{AC}$, 10GHz) across the doped silicon slot PCW, simulated by COMSOL MULTIPHYSICS, with the antenna replaced by perfect conductors. Fig. 3 (b) shows that the voltage drop (>90%) mostly occurs inside the slot. As the RF frequency ($f_{RF}$) increases, the impedance of the slot ($1/(C\omega)$), where C is slot capacitance and $\omega=2\pi f_{RF}$) decreases, and the fraction of the voltage dropped across the slot is reduced due to the finite resistance of the silicon PCW. For this two-level doping condition, a simulation using LUMERICAL DEVICE software shows that the total resistance of our 300 μm-long silicon PCW is 189 Ohms and the slot capacitance is 38.6fF, so the limiting RF frequency bandwidth of the device can be estimated to be $1/(2\pi RC)$=22GHz.

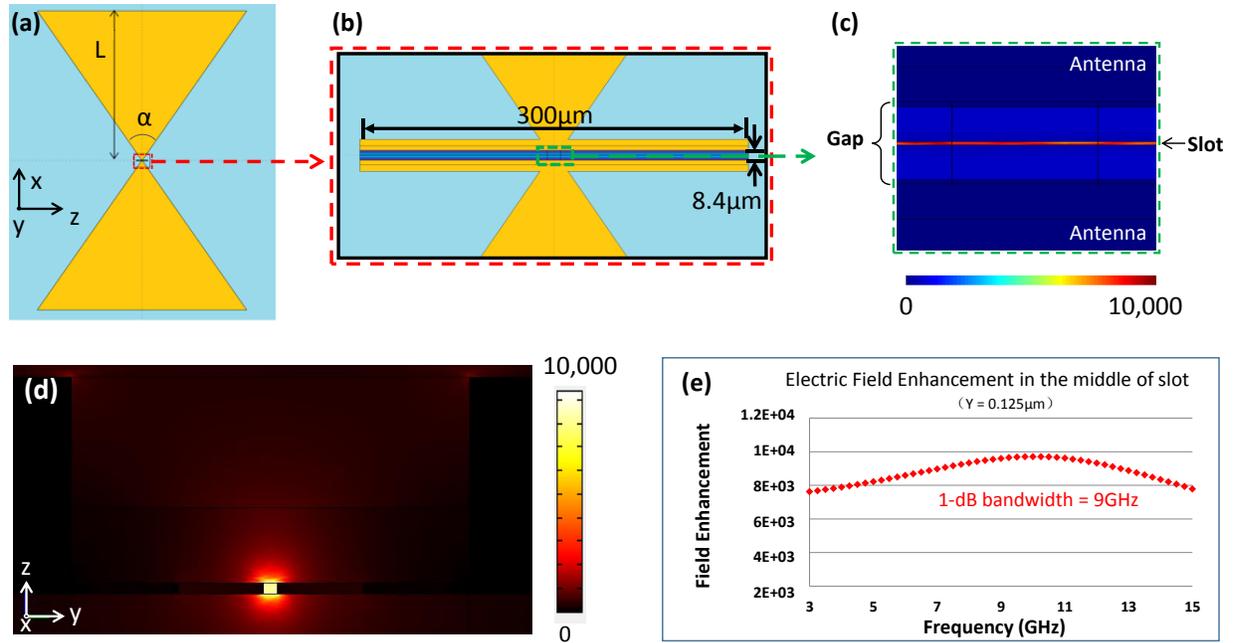

Fig. 4. (a) Schematic top view of the designed bowtie antenna. Arm length, L=3mm, and flare angle, α=60°. (b) Magnified image of the feed gap region in (a). (c) Top view of electric field enhancement distribution inside the feed gap of the antenna. The electric field enhancement distribution is shown inside the EO polymer refilled slot at y=0.125μm, where y=0 corresponds to the horizontal interface between the silicon layer and the buried oxide layer. (d) Cross-sectional view of the electric field enhancement distribution inside the feed gap of the antenna. (e) Variation of the field enhancement factor inside the slot versus incident RF frequency.

As shown in Figs. 4 (a) and (b), the antenna is a conventional bowtie antenna with capacitive extension bars attached to the apex points of the bowtie, in order to obtain an extended area with a strong uniform electric field enhancement. The extension bars have a length of 300μm, which is the same as the length of EO polymer refilled slot PCW. The narrow feed gap width between the two capacitive bars is 8.4μm, for the generation of highly enhanced local electric field under RF illumination. The current on the bowtie antenna surface, induced by incident RF field, charges the feed gap and subsequently establishes this strong electric field in the feed gap [56]. The thickness of the gold film is chosen to be 5μm, which is far beyond the skin depth of gold at the RF frequency of operation. This gold antenna is designed with bow arms on silicon dioxide to avoid the impact of conductive silicon region. In the actual device fabricated on an SOI substrate, the top silicon region everywhere apart from the PCW region is entirely etched away, as shown in Fig. 5 (b), letting the buried oxide layer be directly underneath the bow arms to fit this design. The silicon handle underneath the buried oxide layer is taken into account in the simulation.

Generally, the antenna system can be considered as a typical LC circuit, which is mainly composed of the inductive bowtie metallic arms and capacitive bars filled with EO polymer, giving rise to an LC resonance determined by antenna geometry [57]. In this work, this resonance effect is characterized by field enhancement (FE) factor, defined as the resonant electric field amplitude (inside the slot) divided by the incident electric field amplitude at the specific observation point. With the feed gap width and capacitive bars fixed, the resonant frequency of a bowtie antenna is mainly determined by the length of each bow arm and the flare angle [L and α in Fig. 4 (a)] [58]. This bowtie antenna structure together with the effective-medium approximated silicon RF dielectric constant and conductivity values is used for COMSOL MULTIPHYSICS simulation. With bow arm length L=3mm and flare angle α =60°, the bowtie antenna is optimized with a central resonant frequency at around 10GHz, and a uniform electric field enhancement over the entire feed gap is created. Figs. 4 (b) and (c) show the top view of the local resonant electric field amplitude inside the antenna feed gap at 10GHz. Fig. 4 (d) shows the cross-sectional view of the electric field enhancement distribution inside the feed gap of the antenna. The electric field is mainly confined in the feed gap region, which is similar to the performance of a typical dipole antenna [59]. Additionally, as explained above, the electric field is actually concentrated inside the slot of the silicon PCW, and this increases the FE even further. Fig. 4 (e) shows the FE spectrum from simulations, indicating that the electric field radiation compressed inside the slot of the silicon PCW is enhanced by a maximum factor of ~10,000 at 10GHz, with a 1-dB RF bandwidth over 9GHz. This strongly enhanced RF field directly modulates the optical wave propagating along the EO polymer refilled doped silicon slot PCW which is embedded inside the feed gap. No extra connection lines between the antenna and EO modulator and no external electrical power supply are required [12].

## 3. FABRICATION

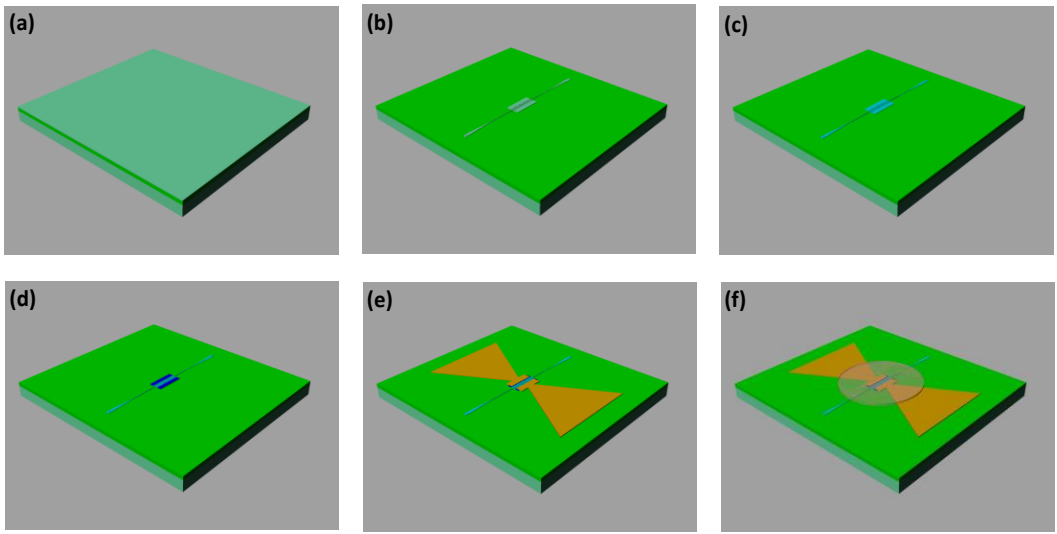

Fig. 5. Fabrication flow. (a) SOI wafer, (b) Silicon photonic waveguide, patterned by electron-beam lithography, RIE, photolithography, and RIE again, (c) 1st ion implantation, (d) 2nd ion implantation, followed by rapid thermal annealing (e) Gold bowtie antenna, patterned by seed layer deposition, photolithography, electroplating, and seed layer removal, (j) Spincoating of EO polymer (indicated by the circle area), followed by vacuum oven baking and EO polymer poling.

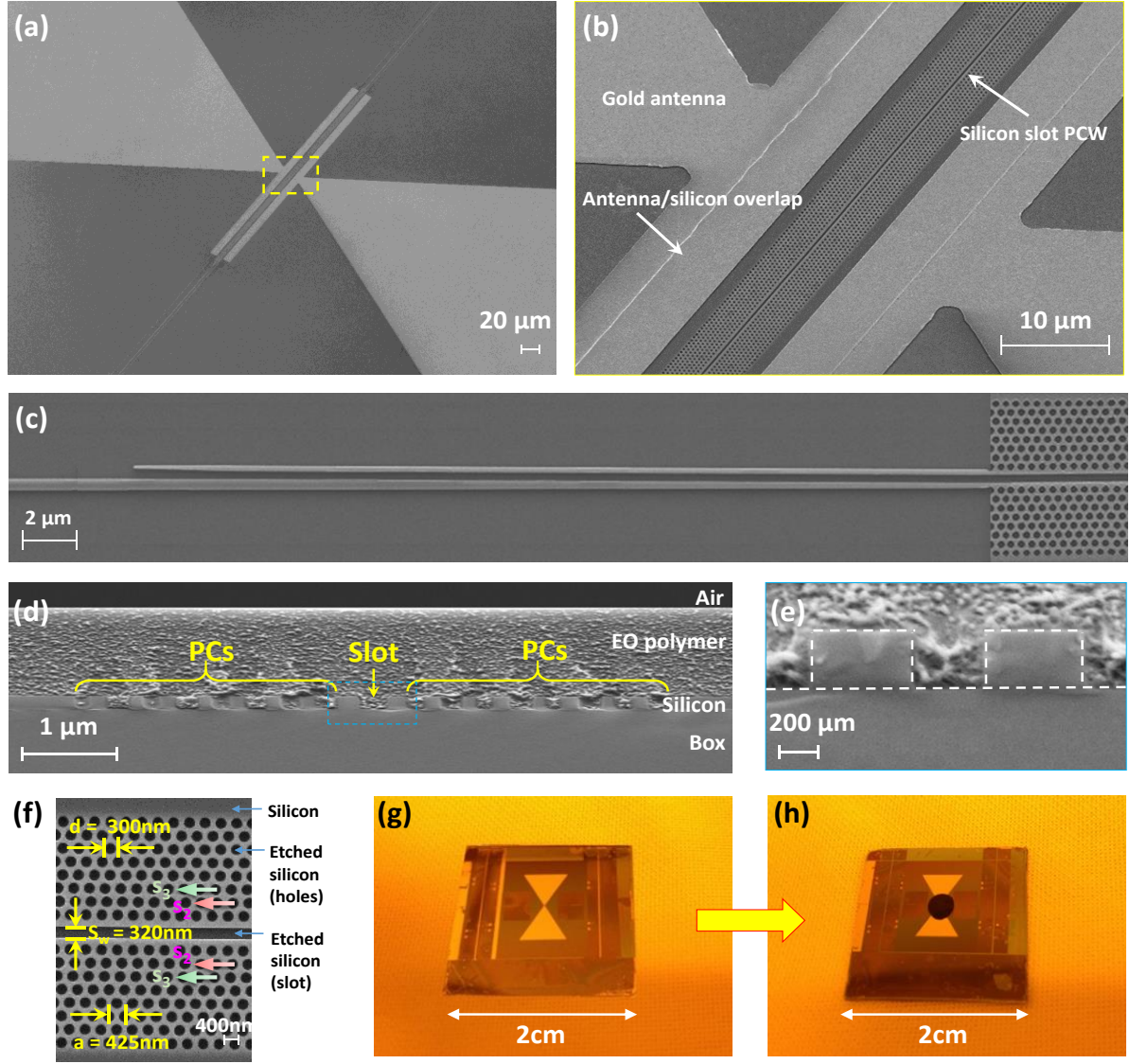

Fig. 6. (a) SEM image of the fabricated device. (b) Magnified SEM image of the yellow rectangular region in (a) showing the slot PCW region and bowtie antenna overlay. (c) SEM image of the strip-to-slot mode converter for efficient coupling between strip waveguide and the 320nm slot PCW. Note: in (a)-(c) the device is not covered by EO polymer, just for better visualization. (d) SEM image of the cross section of the EO polymer refilled silicon slot PCW. PCs: photonic crystals. (e) Magnified SEM image of the blue rectangular area in (d). (f) Top view of the band-engineered slot PCW region. The arrows show the direction of lattice shifting. (g) and (f) The real device before and after EO polymer is spincoated.

The fabrication flow is briefly illustrated in Fig. 5. The fabrication starts with an SOI wafer with a 250nm-thick top silicon and a 3μm-thick buried oxide layer, as shown in Fig. 5 (a). The silicon slot PCW is fabricated using electron-beam lithography and reactive ion etching (RIE). Next, all of the top silicon region, except the area with the slot PCW, is completely removed by photolithography and RIE, as shown in Fig. 5 (b). Then, the silicon slot PCW is first implanted with P+ at an energy of 92keV and a dose of $1.05 \times 10^{12} cm^{-2}$ to reach an ion concentration of $1 \times 10^{17} cm^{-3}$ [Fig. 5 (c)]. Next, the device is patterned by photolithography and implanted with P+ at an energy of 92keV and a dose of $1.05 \times 10^{15} cm^{-2}$ to reach an ion concentration of $1 \times 10^{20} cm^{-3}$ at the outer sides of the silicon rails which will be connected to the bowtie antenna in order to form Ohmic contacts [Fig. 5 (d)]. A rapid thermal annealing at 1000°C for 10min in a nitrogen environment is then performed to annihilate the induced defects and to activate the implanted ions, which also improves the optical performance of the ion-implanted waveguides [60]. Next, a 50nm-thick gold seed layer with a 5nm-thick chromium adhesion buffer is deposited by electron-beam evaporation, and a buffer mask for the bowtie antenna is patterned on a 10 μm-thick AZ9260 photoresist using photolithography. Then, a 5 μm-thick gold film is electroplated by through-mask plating method in a magnetically stirred neutral noncyanide electrolyte (Techni-Gold 25ES) under a constant current of 8mA at the temperature around 50°C. The AZ9260 buffer mask and gold seed layer are finally removed by lift-off and wet etching, as shown in Fig. 5 (e). SEM images of the fabricated device are shown in Fig. 6. The inner sides of the extension bars of the bowtie antenna are connected to the outer sides of silicon rails which are heavily doped for Ohmic contact between the antenna and the PCW, as shown in Fig. 6 (b).

Next, the EO polymer, SEO125, is formulated and infiltrated into the holes and the slot of the silicon PCW region by spincoating, as shown in Figs. 6 (d), (e), (g) and (h), followed by baking at 80°C in vacuum oven for 12hrs. Finally, to activate the EO effect, a poling process is performed [18, 23, 30]. The device is heated up on a hot plate to the EO polymer glass transition temperature of 150°C in a nitrogen atmosphere, and a constant poling electric field of 110V/μm is applied across the EO polymer inside the, as shown in Fig. 7 (a). The randomly oriented chromophore dipoles inside the polymer matrix are then free to rotate and align in the direction of poling electric field. Next the temperature is quickly decreased to room temperature while the constant electric field is still applied, and eventually the chromophores are locked in a uniform direction to form a noncentrosymmetric structure. During this poling process, the leakage current is monitored and remains below 0.66nA, as shown in Fig. 7 (b), corresponding to a low leakage current density of 8.8A/m$^2$. This indicates that the 320nm-wide slot dramatically reduces the leakage current through the silicon/polymer interface compared to sub-100nm slot designs [39, 43].

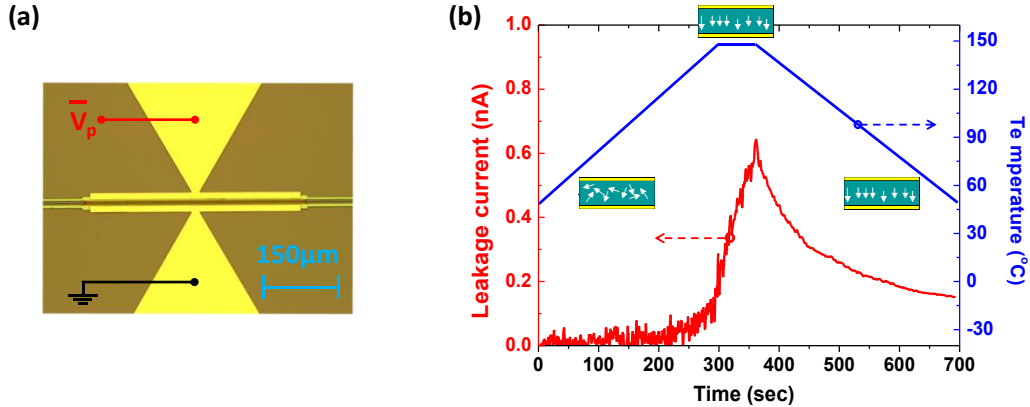

Fig. 7. (a) Microscope images of the fabricated PCW MZI modulator (prior to EO polymer spincoating), overlaid with a schematic of poling configuration. $V_p$, poling voltage. (b) The temperature-dependent leakage current in the EO polymer poling process, overlaid with the variation of chromophore dipole orientations.

## 4. CHARACTERIZATION

In this Characterization Section, first, the bowtie antenna and slot photonic crystal waveguide are tested in Section 4 (A) and Section 4 (B), respectively. After these RF path and optical path are separately demonstrated, then the interactions between the RF wave and the optical wave are demonstrated. Specifically, in Section 4 (C), an EO modulation test is performed, and in Section 4 (D), a high-frequency electromagnetic field sensing is demonstrated.

### A. Bowtie antenna characterization

First, the bowtie antenna on the fabricated sensor, as shown in Fig. 6 (h), is tested as a receiving antenna. RF signal at 10GHz from a vector network analyzer (HP 8510C) is applied to an X-band horn antenna which is mounted on top of the fabricated bowtie antenna. The horn antenna is place sufficiently away in its far field for the assumption of quasi-plane wave to hold. The electromagnetic power that the bowtie antenna receives is measured by a microwave spectrum analyzer (MSA, HP 8560E) via a ground-signal (GS) microprobe (Cascade Microtech ACP40GS500) which contacts the bow arms of the bowtie. Fig. 8 (a) shows the received power of the bowtie antenna. The power response is about 30dB above the noise floor at 10GHz. This simple test demonstrates the functionality of the fabricated bowtie antenna. Considering the reciprocity, the bowtie antenna should also work well as a transmitting antenna.

In order to demonstrate the broadband characteristics of the fabricated bowtie antenna, the vector network analyzer is used to measure the S$_{11}$ parameter (reflection coefficient) of the bowtie antenna. The GS microprobe is used to couple RF power from the vector network analyzer into the bowtie antenna, and the $S_{11}$ parameter over a broad frequency range from 1-16GHz is recorded. Assuming negligible loss, the transmission factor can be inferred from the $S_{11}$ measurements, as shown in Fig. 8 (b), from which a broadband response can be clearly seen. The maximum response occurs at 10GHz, which agrees well with the simulated maximum field enhancement at 10GHz in Fig. 4 (e). The resonance frequency of the bowtie antenna can be tuned by adjusting the bowtie geometry, such as its arm length and glare angle. Details about the dependence of the resonance frequency on bowtie geometry are investigated in Ref. [37]. The broadband characteristics of this bowtie antenna indicates that our sensor can be used to detect the electromagnetic field over a broad frequency bandwidth in the GHz regime.

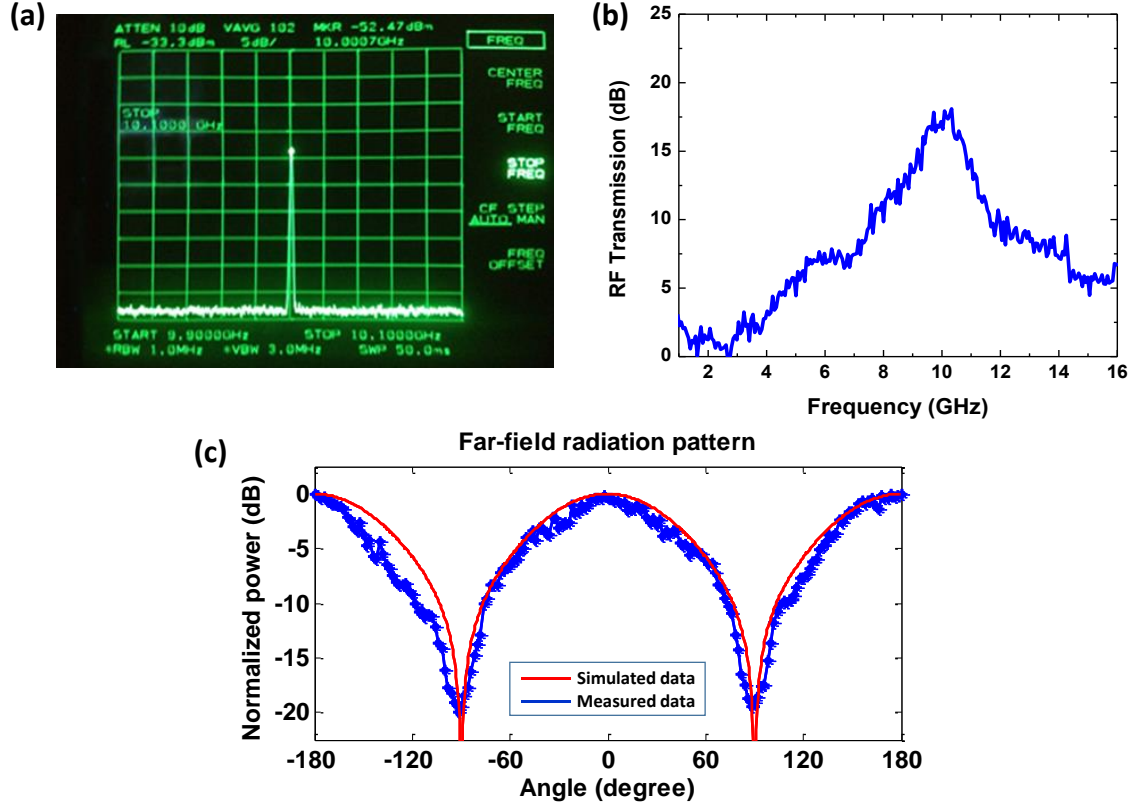

Fig. 8. (a) Measured response power of the bowtie antenna as a receiving antenna at 10GHz. (b) Measured transmission signal of the broadband bowtie antenna. (c) Measured far field radiation pattern (blue) of the bowtie antenna at a frequency of 10GHz. Simulated radiation pattern is also overlaid (red).

In addition, the far field radiation pattern of the bowtie antenna is measured. The bowtie antenna is mounted on a rotational stage and is rotated along the z axis in Fig. 1 (e). RF signal from the vector network analyzer is coupled into the bowtie antenna through a GS microprobe. The frequency of this RF signal is set to 10GHz which is the resonant frequency of the bowtie antenna. A horn antenna is placed 2m away as a receiving antenna in the far field region. The received power is amplified by an RF amplifier and then measured by the microwave spectrum analyzer. The normalized measured power as a function of rotation angle is shown as a blue curve in Fig. 8 (c). Simulated radiation pattern (red curve) is also overlaid in the figure, showing a good match between simulation and experimental results. This measured radiation pattern indicates dipole-type characteristics of our bowtie antenna. The half power beam width is measured to be about 90 degrees. This wide beam width is good for the antenna to detect electromagnetic waves coming from a large range of incident angles.

**B. Optical waveguide characterization**

In order to test the EO polymer refilled silicon slot PCW, light from a broadband amplified spontaneous emission (ASE) source (Thorlabs ASE730) is coupled into and out of the device utilizing an in-house built grating coupler setup

[41, 42]. The optical output signal is observed on an optical spectrum analyzer (OSA, Ando AQ6317B). Fig. 9 shows the measured transmission spectrum of the fabricated EO polymer refilled silicon slot PCW normalized to a reference strip waveguide on the same chip. From the normalized transmission spectrum, a clear band gap with more than 25dB contrast is observed, indicating efficient coupling into the slot PCW [46].

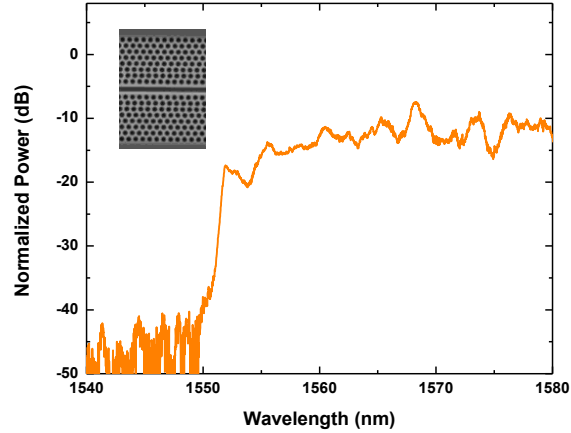

Fig. 9. Measured normalized transmission spectrum of the EO polymer refilled silicon slot PCW. The inset shows an SEM image of the fabricated silicon slot PCW.

## C. Electro-optic modulation experiment

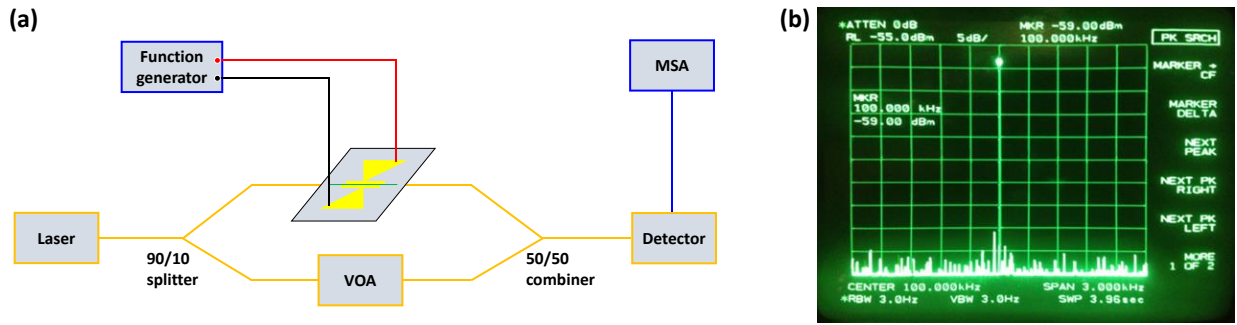

Fig. 10. (a) The schematic of the system for EO modulation experiment. VOA: variable optical attenuator. MSA: microwave spectrum analyzer. (b) The EO modulation response signal as measured on the MSA. (c) EO modulation transfer function.

With the RF path and optical path separately verified, as explained in Section 4 (A) and (B), an EO modulation experiment is performed. An MZI system is formed, using a 90/10 polarization maintaining fiber splitter (Thorlabs L130354603), a 50/50 polarization maintaining fiber combiner (Thorlabs L110313487) and a variable optical attenuator (VOA, Thorlabs VOA50PM-FC), as shown in Fig. 10 (a). A low-frequency modulation test is first performed here to verify the functionality of this MZI system and also to confirm the successful poling of EO polymer. A tunable laser source (Santec ECL200) is used to provide TE-polarized optical input. The optical wavelength is tuned to 1556nm which is within the low-dispersion region of the band-engineered PCW [61]. The laser input is split by a 90/10 splitter, in which 90% of the optical power is coupled into and out of the sensor device through subwavelength grating (SWG) couplers and 10% goes to the external arm with the VOA. Next, the VOA is adjusted until the optical power at the output of the external arm is equal to that coming out of the sensor device. A 50/50 combiner is then used to combine the optical waves from the two arms, so that the phase modulation can be converted into intensity modulation at the output of this MZI system. A sinusoidal RF signal with a peak-to-peak voltage ($V_{pp}$) of 1V at a frequency of 100KHz is generated using a function generator (Agilent 33120A) and directly applied across the two arms of the bowtie antenna. In this case, the two arms work as lumped-element driving electrodes and directly modulate the optical waves propagating in the slot PCW embedded in the feed gap of the bowtie antenna. The modulated output optical signal is converted back to an electrical signal using an amplified photodetector (Thorlabs

PDA10CS), whose power is measured on the microwave spectrum analyzer (MSA), as shown in Fig. 10 (b). The measured response signal indicates that the optical signal is modulated at the same frequency as the input RF signal. When the laser is switched off, this response signal on the MSA disappears, confirming that the signal measured by the MSA originates from the real EO modulation instead of RF cross talk. The EO modulation experiment also demonstrates the successful poling of the EO polymer, as well as the functionality of the MZI system which is then used in the sensing experiment in Section 4 (D).

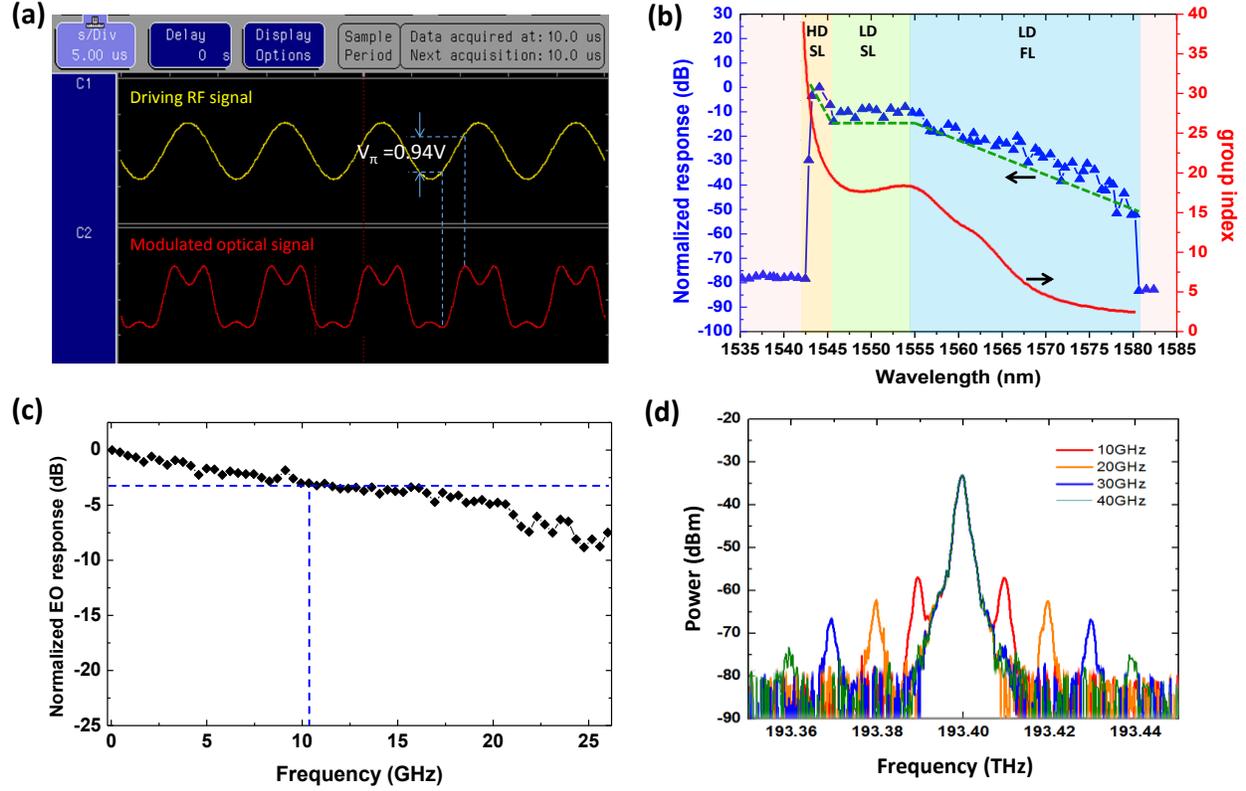

Fig. 11. (a) The measured transfer function at 100kHz. The $V_\pi$ is measured to be 0.94V from over-modulation. (b) Normalized device response v.s. wavelength (at 100KHz). The green dashed line indicates the trend of the response change over different wavelength. The simulated $n_g$ v.s. wavelength is also overlaid. (c) Measured normalized EO response of the modulator as a function of modulation frequency in a small-signal modulation test. The 3-dB bandwidth is measured to be 11GHz. (d) Measured optical transmission spectra of the modulator operating at 10GHz, 20GHz, 30GHz and 40GHz.

Another EO modulation test further demonstrates the high modulation efficiency, wide optical bandwidth, and broad RF bandwidth, in which an integrated MZI modulator is fabricated with the same EO polymer refilled PCW on both arms [39]. The modulator is biased at the 3dB point and driven by a sinusoidal RF wave with a peak-to-peak voltage of $V_d$=1.5V at 100KHz. The modulated output optical signal is detected using an amplified photodetector, and the modulation transfer function is measured by a logic analyzer (HP 1660ES). As shown in Fig. 11 (a), over-modulation is observed on the output optical waveform, in order to measure the half-wave switching voltage ($V_\pi$) of the modulator. The $V_\pi$ is measured to be 0.94V. Thus, the voltage-length product of this modulator achieved is $V_\pi \times L$=0.94V×300μm=0.282V×mm. The effective in-device $r_{33}$ is then calculated to be [22]

$$r_{33,eff} = \frac{\lambda S_w}{n^3 V_\pi \sigma L} = 1230 \text{ pm/V} \qquad (4)$$

where λ=1550nm, $S_w$=320nm, n=1.63, L=300μm, and σ=0.33, which is the highest ever in-device $r_{33}$ ever recorded. This agrees well with the theoretical value in Section 2 (B). Such a record-high $r_{33}$ value originates from the combined effects of a large bulk $r_{33}$ of the EO polymer material, an improved poling efficiency achieved via widening the slot width (320nm), and the slow-light enhancement in the silicon PCW.

Next, a small signal modulation test is done at $V_{pp}<1V$ over a range of wavelength from 1535nm to 1582nm, while all other testing conditions remain the same. The measured normalized modulation response as a function of wavelength is plotted in Fig. 4 (b), and the simulated group index ($n_g$) is also overlaid. It can be seen that the response is almost flat in the low-dispersion slow-light region (wavelength from 1546.5nm to 1554.5nm), because the slot PCW is band-engineered to have a nearly constant $n_g$ in this wavelength range, as shown in Fig. 2 (e). Therefore, our device will be operated within this low dispersion wavelength range.

The RF bandwidth is measured in a high-frequency small-signal modulation test ($V_d \approx 0.61V$). RF driving signal is provided by the vector network analyzer and applied onto the electrodes via the microprobe. The modulated optical signal is amplified by an erbium doped fiber amplifier (EDFA) and received by a high-speed avalanche photodetector (Discovery Semiconductors, DSC40S), and then the received power is measured using the microwave spectrum analyzer. The measured EO response of the device as a function of modulation frequency is shown in Fig. 10 (c), from which a 3-dB bandwidth of 11GHz is measured. Next, another measurement is performed to demonstrate the high-frequency modulation using a sideband detection technique [34, 62-65]. The transmission spectrum of the modulator is measure by the optical spectrum analyzer. Figure 10 (d) shows overlaid transmission spectra of the optical modulator driven at 10GHz, 20GHz, 30GHz and 40GHz, indicating the achievable EO modulation up to 40GHz.

### D. Electromagnetic field sensing experiment

Finally, the electromagnetic field sensing experiment is carried out. The schematic of the experimental system setup is shown in Fig. 12 (a). A sweep oscillator (8620C HP, 2-8.4GHz) is used as a high-frequency RF source to provide an RF signal at a frequency of 8.4GHz. This RF signal is applied to an X-band horn antenna with a gain of 6dB. In this case, the horn antenna works as a transmitting antenna to generate electromagnetic field in free space, and the bowtie antenna in our device works as a receiving antenna to detect the electromagnetic field impinging upon it. The horn antenna is placed at a distance of 30cm vertically over the sensor device, which is beyond the far-field distance of the horn antenna so that the electromagnetic waves radiating on the sensor can be treated as plane waves in the following experimental data analysis. The laser wavelength is tuned to 1556nm which is within the low-dispersion region of the band-engineered PCW. The high-speed avalanche photodetector is used to detect the modulated optical signal at high frequency, and the microwave spectrum analyzer is used to measure the sensing signal.

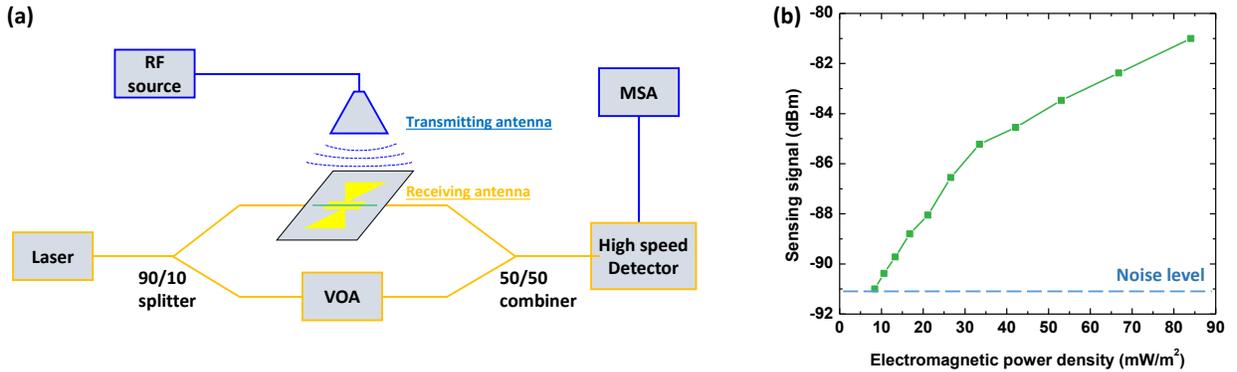

Fig. 12. The measured sensing signal at 8.4GHz as a function of the electromagnetic power density at the position of sensor device.

To characterize the sensitivity of this sensor in terms of electromagnetic power density (or electric field magnitude), the RF power applied on the horn antenna is varied. The corresponding variation in the electromagnetic power density radiating to the position of sensor device is calculated based on Eq. 4 [66].

$$S_{avg} = \frac{G_t P_t}{4\pi R^2} \quad (5)$$

where $S_{avg}$ is the average Poynting vector (electromagnetic power density, unit: mW/m$^2$), $G_t$=6dB is the gain of the transmitting horn antenna, $P_t$ is the input RF power applied on the horn antenna, R=30cm is the distance between the horn antenna and the sensor device. The measured sensing signal as a function of electromagnetic power density is plotted in Fig. 12. It can be seen that the sensing signal decreases as the electromagnetic power density decreases. When the electromagnetic power density decreases to 8.4mW/m$^2$ (equivalent to the input RF power of 2dBm applied on the horn antenna) at 8.4GHz, the sensing signal is below the noise level. Based on Eq. 6, this minimum detectable electromagnetic power density (8.4mW/m$^2$) is used to estimate the minimum detectable electric field amplitude ($|E|$) as 2.5V/m at 8.4GHz, considering the electromagnetic field has a predominantly plane-wave character within the far-field region of the horn antenna [67].

$$|E| = \sqrt{\frac{2S_{avg}}{\varepsilon_0 \varepsilon_r c}} = 2.5 \text{V/m} \qquad (6)$$

where $\varepsilon_0 = 8.85 \times 10^{-12}$F/m is the vacuum dielectric constant, $\varepsilon_r \approx 1$ is the dielectric constant of air, $c = 3 \times 10^8$m/s is the speed of light. Given the FE, for this incident field, the electric field inside the slot is about $2.5 \times 10^4$V/m. The measurement of the maximum detectable electric field is limited by the upper limit of the output power of the RF source. In practice, the maximum detectable electric field is expected to be very high and is determined by the breakdown electric field of the EO polymer material ($>1 \times 10^8$V/m). Using the sensitivity defined in [14], our electromagnetic field sensor has an ultra-high sensitivity of 0.000027V/m Hz$^{-1/2}$ ever demonstrated. As for the detectable electromagnetic field frequency, our sensor has the potential to detect the electromagnetic field over a broad bandwidth, because the bowtie antenna has been demonstrated for broadband operation in GHz frequency regime in section IV (A).

## 5. PROPOSED FUTURE WORK

**A. Terahertz wave sensor based on electro-optic polymer filled plasmonic slot waveguide**

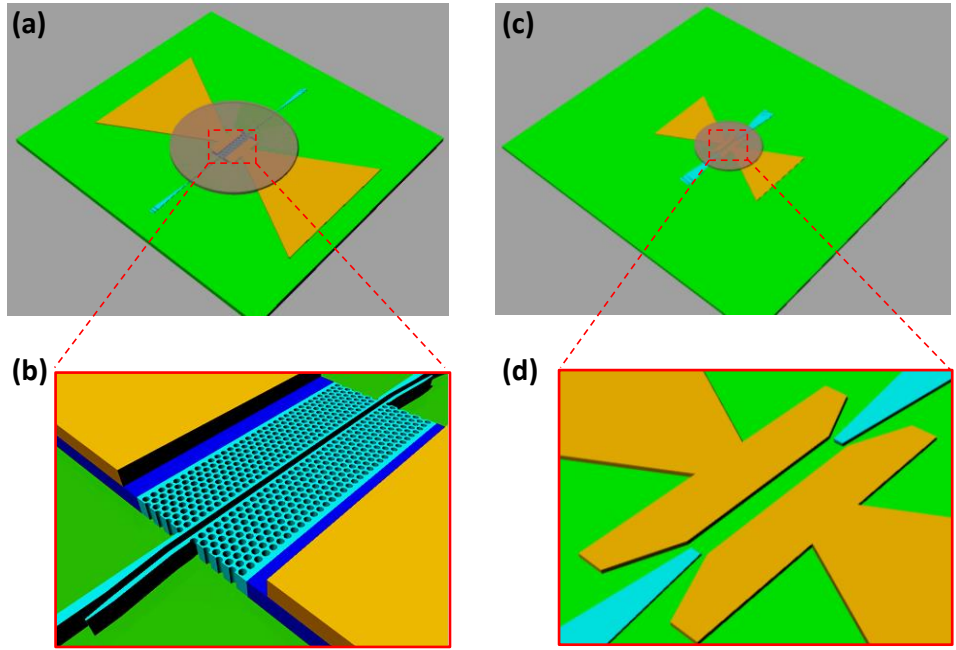

Fig. 13. Evolution from a X-band electromagnetic wave sensor to Terahertz wave sensor. (a) A schematic view of the electromagnetic field sensor consisting of an EO polymer refilled silicon slot PCW phase modulator and a bowtie antenna. (b) Magnified image of slot PCW. (c) A schematic view of the electromagnetic field sensor consisting of an EO polymer refilled plasmonic slot waveguide phase modulator and a bowtie antenna. (d) Magnified image of plasmonic slot waveguide.

In previous sections, we have demonstrated an integrated photonic electromagnetic field sensor to cover the X band of the electromagnetic spectrum (8-12GHz). The key parts of this X-band photonic electromagnetic field sensor are shown schematically in Figs. 13 (a) and (b), including an EO polymer refilled silicon slot PCW phase modulator and a gold bowtie antenna [22]. The resonance frequency is targeted at 10GHz, which is the central frequency of X band. In our next stage of research, we are trying to develop an advanced version of this sensor to cover wide electromagnetic spectrum up to 10THz, including the range of microwave, millimeter wave and even terahertz wave. In this section, we will talk about the proposal of a Terahertz wave sensor.

To make our sensor to cover a large range up to THz frequency regime, the device is modified, as shown in Figs. 13 (c) and (d). First, the arm length of the bowtie antenna is scaled to provide a broad bandwidth with central resonance frequency in Terahertz range. Figure 14 (a) shows the simulated S11 parameter of the bowtie antenna, with a resonance frequency around 8THz, and Fig.14 (b) shows the simulated normalized electric field distribution on bowtie antenna at 8THz with a strong near-field enhancement inside the feed gap. Second, the silicon slot PCW is removed and the gap between the two extended bars are narrowed down to 250nm to form a plasmonic slot waveguide. The metal slot is filled with EO polymer, and optical modulations can be achieved in this plasmonic slot waveguide as demonstrated recently in Ref. [64]. Figure 14 (c) shows the simulated mode profile of the plasmonic slot waveguide, from which it can be seen that most of optical power is confined inside the slot. The modulation electric field is also concentrated in the slot, as shown by the simulation results in Fig. 14 (d), leading to a very large overlap between optical mode and modulation field and thus a very high modulation efficiency. In addition, a silicon strip waveguide to plasmonic slot waveguide mode converter is designed to efficiently couple light into the surface plasmonic mode [68]. The working principle of this Terahertz wave sensor is based on the electric field enhancement by the bowtie antenna and the electro-optic modulation inside the plasmonic slot waveguide, similar to the X-band electromagnetic wave sensor. By measuring the modulated optical signal, an incident Terahertz wave can be detected through optical means.

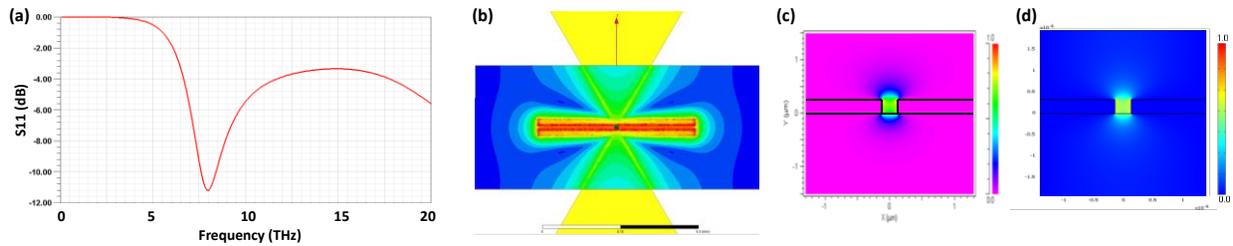

Fig. 14. (a) Simulated S11 parameter over a frequency range 1-20THz. (b) Simulated normalized electric field distribution on bowtie antenna at 8THz. (c) Simulated plasmonic slot waveguide mode profile by Rsoft (d) Simulated RF electric field profile on the plasmonic slot waveguide by COMSOL.

There are several important benefits of this plasmonic-based Terahertz wave sensor. (1) Due to the removal of silicon PCW and the high conductivity of metals, the RC time delay is significantly reduced, enabling a very broad frequency bandwidth. The resonance of the bowtie antenna can be engineered to cover a frequency range from 1GHz to 10THz. Based on simulation results as shown in Fig. 15 (a), the normalized electric field dropped across the metal slot is almost constant from the frequency of 1GHz to 10THz. As a comparison, for silicon PCW slot, the normalized electric field across the decreases as the frequency increases due to the finite conductivity of the silicon. Its 3-dB bandwidth is about 30GHz, which limits the detectable frequency band of the sensor. Figures 15 (b) and (c) shows the electric field potential distribution on the plasmonic slot waveguide and the silicon slot PCW at frequencies up to 10THz. It can be seen that the most of the modulation field is still efficiently dropped across the plasmonic slot where EO polymer is filled. (2) The electric field enhancement factor is increased when the bowtie feed gap is reduced, so the 250nm-wide metal slot can provide higher local electric field enhancement which improve the sensitivity of this sensor. A 2X improvement in sensitivity is expected compared to 320nm-wide silicon slot PCW. The 250nm metal gap is not optimized yet, and the sensitivity can be further improved by reducing the metal slot width. (3) The large overlap between the modulation electric field and the surface plasmonic mode enables more efficient modulation, which can reduce the required interaction length. Although plasmonic waveguides have higher propagation loss per unit length compared to regular waveguides, the short device length is helpful in reducing the total propagation loss. In Ref. [64], an interaction length of only 29μm and a total insertion loss of 12dB are demonstrated. Based on the simulated complex effective index achieved in Fig. 15 (c), the propagation loss of our un-optimized plasmonic slot waveguide is 0.320dB/μm. we expect the total insertion loss of our device can be controlled below 10dB, with the optimization of the plasmonic slot waveguide, the selection of metal with lower propagation loss (i.e. copper to replace gold), as well as improved fabrication quality provided by CMOS foundry. (4) By properly designing the plasmonic slot waveguide and engineering its dispersion diagram, a large group index can be achieved in the surface plasmonic mode and this slow-light effect can be used to enhance the modulation efficiency and the detection sensitivity. (5) Some existing electronic-based THz detector requires low temperature to reduce electron noise, so it required a cooling process such as liquid Nitrogen flow; on the contrary, our photonic (plasmonic) detection approach has good noise immunity. Our device can also reduce the impact of perturbing fields, since it is based on an optical modulation technique.

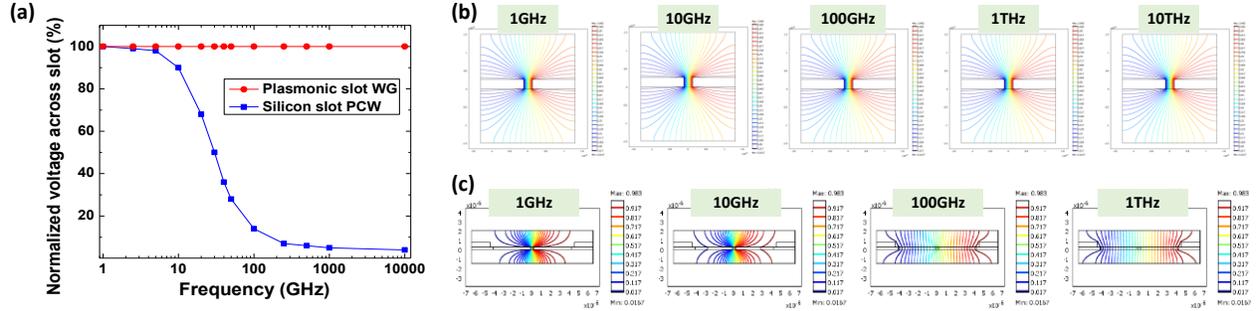

Fig. 15. (a) Simulated normalized voltage drop across the silicon slot PCW and the plasmonic slot waveguide. (b) Simulated electric potential distribution on the plasmonic slot waveguide, at 1GHz, 10GHz, 100GHz, and 1THz, respectively. (c) Simulated electric potential distribution on the silicon slot PCW, at 1GHz, 10GHz, 100GHz, respectively.

To the best of our knowledge, this will be the first Terahertz wave sensor based on electro-plasmonic modulation. The working bandwidth of this sensor can be tuned by modifying the bowtie antenna geometry, in order to enable more potential applications in other frequency ranges such as microwave detection, millimeter wave imaging, and even possibly light trapping for photovoltaics.

### B. Fully packaged and tailgated electromagnetic wave sensor

In our future work, we will deliver a pigtailed electromagnetic wave sensor. We envision that the fully packaged device will look like the schematic shown in Figs. 16. In order to prepare a robust package for the device, we will use the technique of angle-polished fiber-grating coupler proposed in Ref. [69]. A schematic of the angle-polished fiber-grating coupler is illustrated in Fig.1 (c) of Ref. [69]. In this technique, the fiber is placed horizontally and fixed onto the surface of chip using a UV curable epoxy. The epoxy is dropped away from the grating and the polished surface. We are in the process of characterizing angle-polished fibers from OE Land Inc. A polished facet at $\theta_f=41.6°$ provides a final $\theta_{out}=10°$ out-going beam, while the beam hits the polished fact at an angle larger than the total-internal-reflection angle and is thoroughly reflected. A subwavelength grating coupler will be designed for optimum coupling with this angle-polished fiber. We will also investigate if the reflectivity can be enhanced by coating the angled-polished facet by a metal. Holding horizontal fibers in place by epoxy will let us avoid the grating coupling testing setup and it will be significantly easier to bring the horizontal fibers close to each other in order to decrease the length of the passive waveguides. In addition, we will also examine the packaging and reliability issues, such as the tolerance to the fiber tip displacement and rotation, effects of the packaging on the sensor and modulator applications, and the mechanical durability of the couplers.

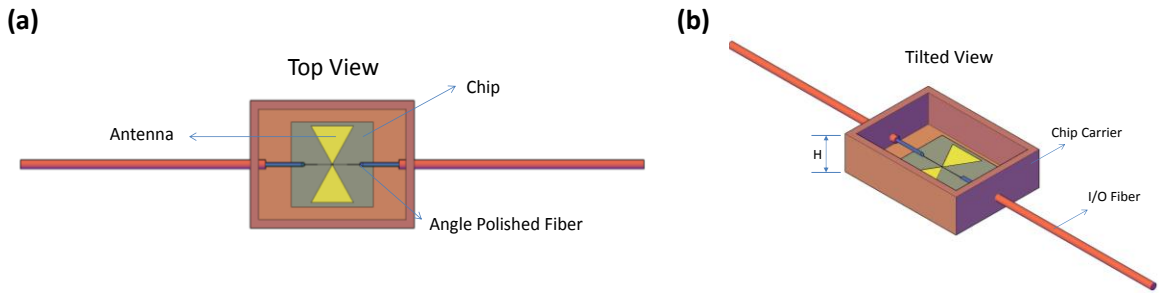

Fig. 16. Schematics of the fully packaged and tailgated electromagnetic wave sensor. (a) Top view. (b) Titled view.

## 6. CONCLUSION

We design, fabricate and experimentally demonstrate a compact and sensitive integrated photonic electromagnetic field sensor based on EO polymer refilled silicon slot PCW coupled with a miniaturized bowtie antenna. The bowtie antenna is used as receiving antenna, poling electrodes and driving electrodes. The bowtie antenna with doped silicon slot PCW embedded inside its feed gap is demonstrated to have broadband characteristics with a resonance frequency at 10GHz and a large beam width of 90 degrees. The modulator is experimentally demonstrated with a record-high effective in-device EO modulation efficiency of $r_{33}$=1230pm/V. Modulation response up to 40GHz is measured, with a 3-dB bandwidth of 11GHz. An 8nm-wide low-dispersion spectrum range is demonstrated, and this makes our device insensitive to the variation of wavelength and temperature, which is much better than non-band-engineered PCWs [45, 70] and ring resonators [4, 71-73]. Therefore large EO coefficient polymer, Slow-light effects in the PCW, as well as broadband electric field enhancement provided by the bowtie antenna, are combined together to enhance the EO modulation efficiency, leading to a very high sensitivity. The minimum detectable electromagnetic power density is demonstrated to be 8.4mW/m$^2$ at 8.4GHz, corresponding to the minimum electric field amplitude of 2.5V/m and the ultra-high sensitivity of 0.000027V/m Hz$^{-1/2}$ [14]. To the best of our knowledge, this is the first silicon-organic hybrid device and also the first PCW device used for the photonic detection of electromagnetic waves. Furthermore, we propose some future work, including a Teraherz wave sensor based on antenna-coupled electro-optic polymer filled plasmonic slot waveguide, as well as a fully packaged and tailgated device.

## ACKNOWLEDGEMENT

The authors would like to acknowledge the Air Force Research Laboratory (AFRL) for supporting this work under the Small Business Technology Transfer Research (STTR) program (Grant No. FA8650-12-M-5131) monitored by Dr. Robert Nelson and Dr. Charles Lee.